# Effect of Cavitation on Velocity in the Near-field of a Diesel Nozzle

H. Purwar,[1,*] K. Lounnaci,[1] S. Idlahcen,[1] C. Rozé,[1] J.-B. Blaisot,[1] L. Méès,[2] M. Michard[2]
[1]CORIA UMR-6614, Normandie Université, CNRS, Université et INSA de Rouen, Avenue de l'Université, 76800 Saint Etienne du Rouvray, France
[2]CNRS, UMR 5509 LMFA, École Centrale de Lyon, 36 avenue Guy de Collongue, 69134 Écully, France

**Abstract**
The entire process of atomization of the fuel in an internal combustion engine plays a very important role in determining the overall efficiency of these engines. A good atomization process could help the fuel to mix with the air properly leading to its efficient combustion, thereby reducing the emitted pollutants as well. The recent trend followed by the engineers focused on designing fuel injectors for more efficient atomization is to increase the atomization pressure while decreasing the nozzle orifice diameter. A consequence of this is the development of cavitation (formation of vapor cavities or bubbles in the liquid) inside the injector close to the nozzle. The main reason behind this is the sudden changes in the pressure inside the injector and these cavities or bubbles are usually formed where the pressure is relatively low.

This work mainly focuses on studying the formation of cavitation and its effect on the velocity of the spray in the near nozzle region using asymmetrical transparent nozzle equipped with a needle lift sensor with nozzle diameter of 0.35 mm at 300 bar of injection pressure.

The experiment consists in recording of several image-pairs, which are separated by about 300 ns, capturing the dynamics of the spray, a few millimeters from the nozzle in the direction of the flow. These image-pairs are then used to compute the velocity from the displacement of the liquid structures and ligaments by correlating the first image with the second. About 200 of such velocity graphs are then averaged to obtain a velocity map and is compared with the similar average velocity maps obtained at different times from the start of the injection. The angular spread of the spray from each of these images is calculated as well. The images showing cavitation inside the injector are also recorded at these same instants of time so as to understand the effects of cavitation on the velocity and angular spread of the spray close to the nozzle.

**Introduction**

A common trend followed by the automobile industry in order to increase the overall efficiency of the internal combustion diesel engines thereby also reducing the emitted pollutants, has been to increase the injection pressures while decreasing the nozzle orifice diameter. This has led to modern diesel injectors that operate at an injection pressure as high as 2000 bar with orifice diameter of about one hundred micrometers. As a result, there is a huge pressure variation inside these injectors, which is one of the main reasons for cavitation. In most diesel injectors such high injection pressures are achieved by reducing the outlet diameter and due to this there is a sudden change in the flow velocity of the fuel, thereby reducing the relative pressure inside the injector, a result from the Bernoulli's principle. Cavitation occurs when the relative pressure drops below the vapor pressure of the fuel resulting in the formation of vapor pockets or cavities near the nozzle outlet.

A lot of efforts have been made in understanding the role of cavitation in the atomization of the fuel sprays [1-7]. It is now believed that cavitation up to an extent favors atomization [1,2] but it depends on a lot of factors not yet understood very well, like internal geometry of the injector, surface smoothness, fuel properties, etc. On the other hand, it has also been shown that supercavitation have negative effects on atomization [4,5]. In fact cavitation could even reduce the lifetime of the fuel injectors because of wear and tear due to the sudden collapse of the cavitation bubbles inside the fuel injectors. However, the same is shown to improve atomization as shown in experiments on axisymmetric nozzles [6].

This work focuses on investigating the impact of cavitation on the macroscopic spray properties, like velocity and spray cone angle for the high-pressure diesel sprays in the near nozzle region using an asymmetrical transparent nozzle for different injection needle lifts.

**Transparent Nozzle Design**

A schematic of the non-symmetrical transparent nozzle used to study the effect of cavitation on the spray velocity and cone angle is shown in Figure 1. A standard nozzle was trimmed at its very end (just above the orifice, under the lift seat) and the transparent tip made of poly-(methyl methacrylate) (PMMA) was fixed using an intermediary metal part. The cylindrical sac volume ends in hemispherical shape and a single orifice which is off-set with respect to the sac volume axis. This asymmetrical configuration aims to obtain an asymmetrical development of cavitation allowing a better visualization of the cavities and liquid-vapor interfaces.

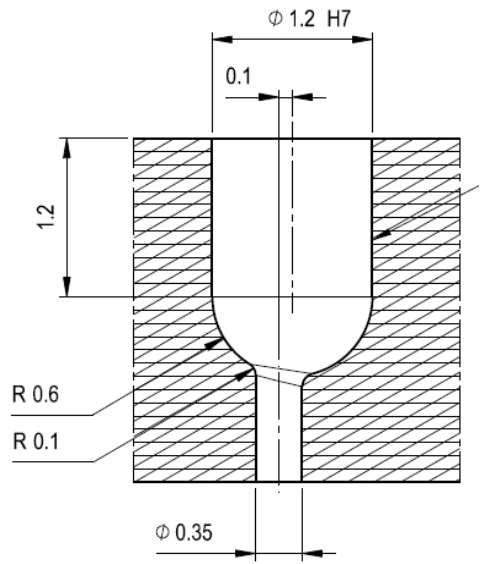

**Figure 1: Schematic of the transparent nozzle made of PMMA.**

**Experimental Design**

Two separate experiments with very similar conditions were designed for the measurement of cavitation and near field velocity of the diesel spray for different injector needle lifts, the details of which are discussed below.

*Cavitation measurements*

A backlit imaging setup was used for recoding cavitation images inside the transparent injector. A double-pulsed ND: YAG laser source (LitronLasers Nano S) illuminated a fluorescent sheet and 12 ns pulses of incoherent light centred at $\lambda = 600$ nm were obtained, as described in ref. [8]. Images of the internal flow were recorded using a CCD camera (Jai TM-4200CL) along with a long distance microscope (QM100, Questar) used to adjust the optical magnification. For these measurements, both the laser and the camera ran at a relatively low repetition rate and as a consequence of this only one image is recorded for each injection. The injection process is then described from several injections by varying the delay between injection trigger signals and recording time, i.e. camera and laser trigger signals.

The liquid used for the cavitation measurements was chosen to have approximately the same refractive index as of PMMA to reduce image distortion. It is mainly composed of a diesel-like liquid (Castrol calibration oil ISO 4113) mixed with a high refractive index liquid.

*Velocity measurements*

The spray velocity in the near nozzle region is estimated by correlating two successive spray images recorded within a very short duration of time – enough for the spray structures and ligaments to be displaced slightly without altering their shapes and sizes significantly. This was achieved by illuminating the fuel sprays with an ultra-short double-pulsed light source, pulses short enough to freeze the motion of the fast moving spray at the detector and a double-frame camera to record high-resolution images of the fuel spray. The time between the two pulses was chosen according to the limitation of the camera and it was made certain that within this short duration the spray is only displaced slightly so that the two captured frames could be correlated to find the displacement of the spray structures and thereafter their velocity.

A schematic of the optical setup for obtaining high-resolution backlit images of the fuel spray is shown in Figure 2. The two successive femtosecond laser pulses – pulse width ~100 fs, wavelength $\lambda = 800$ nm – were generated using two regenerative amplifiers (Libra, Coherent) seeded with a common titanium-sapphire mode-locked laser (Vitesse, Coherent). The delay between the two pulses is adjustable and was set to 314 ns and is limited by the camera's frame rate. These incoming fs laser pulses were separated using a 50:50 beam splitter and were used to obtain spray images for two orthogonal viewing angles as is shown in Figure 2. The dual-frame PIV cameras used for these measurements were Cam1: Imager sCMOS and Cam2: Imager pro X both from LaVision. The entire system, including laser pulses, spray injection and cameras was synchronized using digital delay generators (DG-645) from Stanford Research Systems, operating at a frequency of 1 Hz (limited due to spray injection conditions). The fuel spray from the transparent nozzle was injected at 300 bar inside a closed spray chamber maintained at standard conditions of pressure and temperature. Pure Castrol calibration oil ISO 4113 was used for recording the spray images. Its density and viscosity are mentioned in Table 1 below.

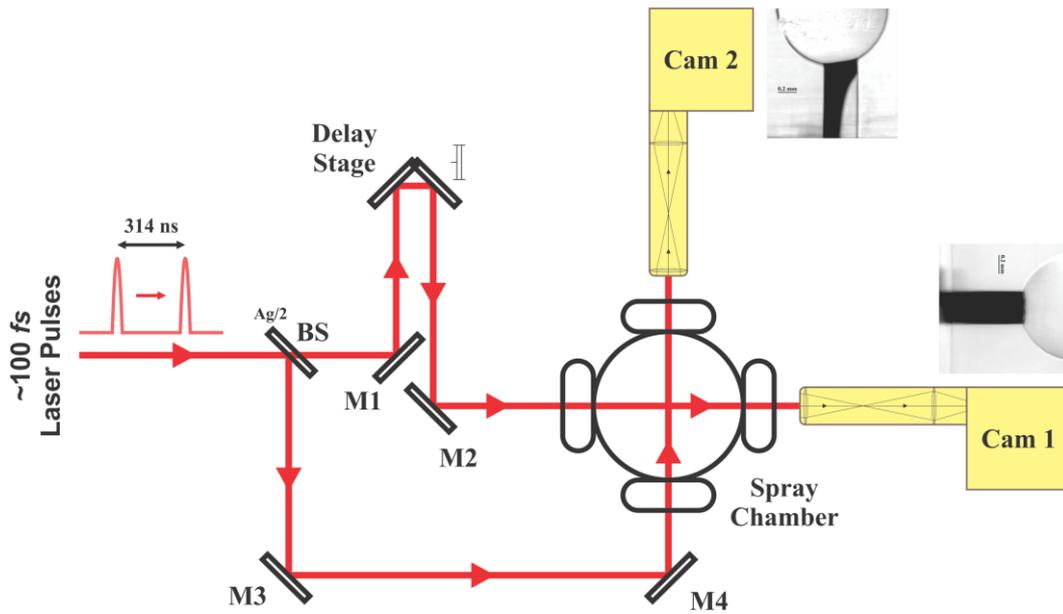

**Figure 2: Schematic of the optical setup for recording high-resolution backlit spray images for two orthogonal viewing angles. Note that Cam 1 and Cam 2 do not record the cavitation images as shown in the schematic but these cavitation images are shown to give an idea of the orientation of the injector nozzle inside the spray chamber.**

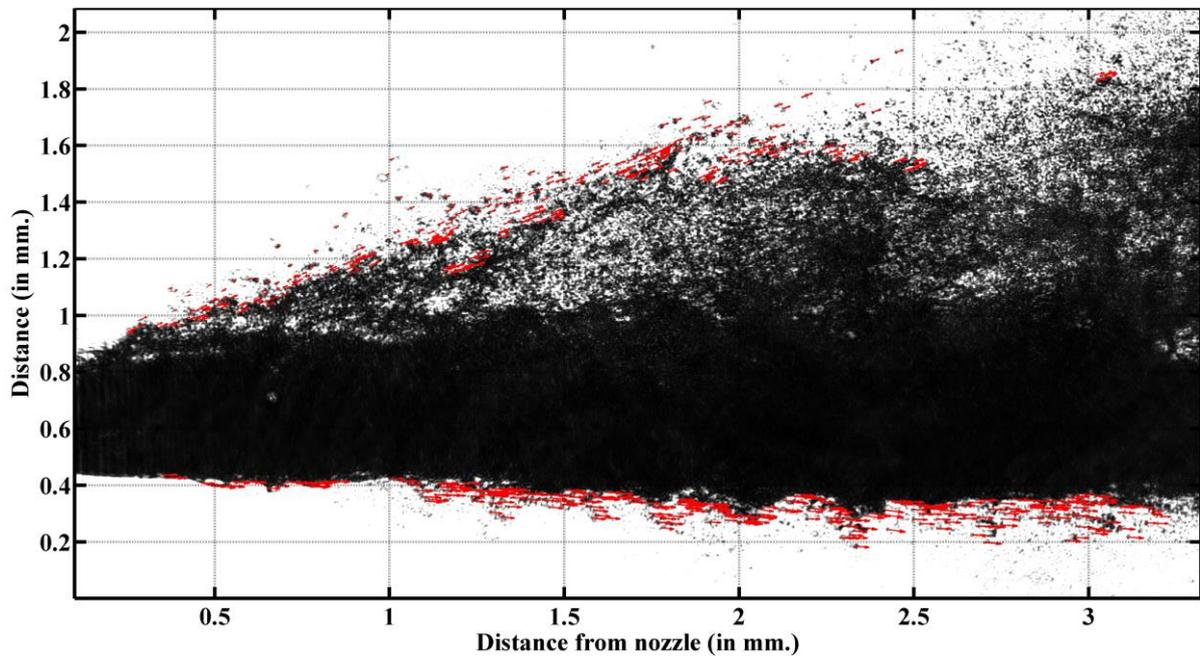

**Figure 3: Sample spray image (first frame) recorded by Cam 2 when 87% of the injector's needle is lifted and the corresponding displacement vectors traced using cross-correlation based approach.**

**Table 1: Physical properties of the diesel-like liquid used for recording spray images for velocity estimation.**

| Liquid | Density (g/ml) | Viscosity (mm$^2$/s) |
|---|---|---|
| Castrol calibration oil ISO 4113 | 0.825 | 2.53 |

The recorded spray image-pairs were normalized and subjected to cross-correlation based velocity computation [9]. A small section of the first image containing the edge of the spray is chosen and then the similar structures and ligaments are searched in the second image. The displacement of the entire chosen section is then esti-mated from the maximum value of the normalized cross-correlation. A sample spray image (first frame) and the corresponding displacement vectors are shown in Figure 3. Note that usually a very small section on the first image of the image-pair is chosen and hence the estimated value of displacement is justifiable. The time for this displacement is the time between the two laser pulses (314 ns). Using this information the velocity of the liquid structures and ligaments visible on the images is estimated.

*Needle lift measurement*

A non-contact, eddy current based displacement measurement system NLS3181 (sensor LS04(04)) from Micro-epsilon for targets made of ferromagnetic electrically conductive materials was used for injection needle lift measurement. High-frequency alternating current flows through a coil cast in a sensor casing. The electromagnetic coil field induces eddy currents in the conductive target thus changing the AC resistance of the coil. This change in the impedance is interrupted by demodulation electronics, which generates an electrical signal proportional to the distance of the target from the sensor.

The cavitation and spray images for velocity estimation were recorded for different injector needle lifts by shifting the laser position with respect to it. The laser positions were shifted using the delay generator itself and were measured using a photo-diode. The signals from the photo-diode and the needle lift sensor were recorded simultaneously using a high-speed data acquisition device (DT9836) from Data Translation.

**Results and Discussion**

In order to understand the role of cavitation on the behavior of the spray, backlit cavitation and spray images were recorded in the near nozzle region for different injector needle lifts by doing two separate experiments with very similar conditions. The injector needle lift measurements and the positions of the laser pulses are shown in Figure 4. The green curves correspond to the cavitation measurements and the red curves correspond to the measurement of the spray images used for velocity and spray cone angle estimation. The laser pulses were shifted precisely using the digital delay generators. The arrival time of the laser pulses used for the measurement of the spray images (in red) and the closest available cavitation measurements (in green) are highlighted in the figure and the corresponding injector needle lift percentages are shown in Table 2.

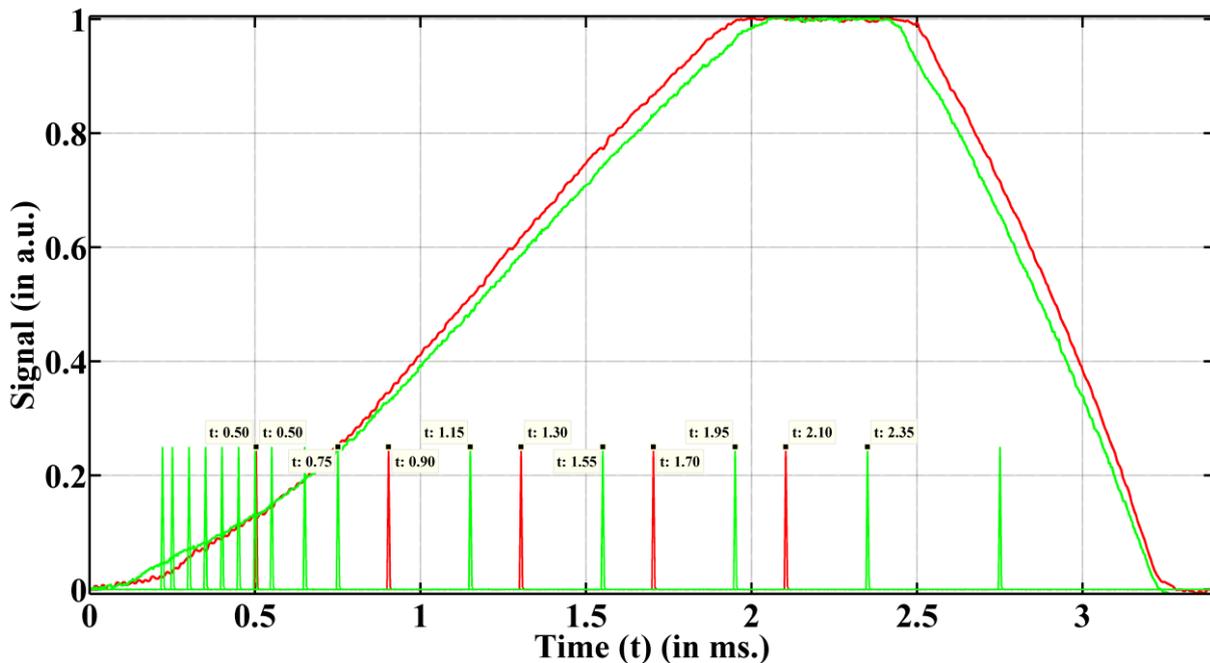

**Figure 4: Injector needle lift signal and corresponding positions of the laser pulses used for cavitation (green) and spray (red) imaging.**

**Table 2: Laser positions for cavitation measurements and recording spray images and the corresponding injector needle lifts.**

| For cavitation imaging | | For spray imaging | |
|---|---|---|---|
| Laser positions (ms.) | Needle lifts (%) | Laser positions (ms.) | Needle lifts (%) |
| 0.50 | 13.2 | 0.50 | 12.9 |
| 0.75 | 24.3 | 0.90 | 34.4 |
| 1.15 | 48.5 | 1.30 | 61.7 |
| 1.55 | 74.0 | 1.70 | 86.7 |
| 1.95 | 96.4 | 2.10 | 99.8 |
| 2.35 | 100 | | |

Figure 5 shows the average cavitation images (sample size 25 images) for the laser positions mentioned in Table 2, where the RI matched diesel-like liquid is injected with an injection pressure of 300 bar. As a consequence of the RI matching, refraction of light at the nozzle/liquid boundaries is not visible and the liquid phase is bright in the shown images whereas vapor or cavitation pockets are dark due to the interaction of light with the liquid/gas interfaces. For the spray images the liquid spray is dark and the surrounding air is bright. The average cavitation images can also be interpreted as the probability of occurrence of vapor pockets inside the transparent nozzle. Hence, more black is the pixel, higher is the probability of having cavitation in that region. The variation of total intensity count from these average cavitation images is shown in Figure 6 versus time after the start of injection, measurement point corresponding to all the laser positions shown in green in Figure 4. Note that average images were inverted (in grayscale) before calculating the total cavitation intensity so that, higher is the intensity count, higher is the probability of cavitation inside the injector nozzle in Figure 6. Green markers indicate the time with respect to the injector needle lift at which the cavitation images shown in Figure 5 were recorded and red markers indicate the time at which spray images were recorded for velocity and cone angle estimation.

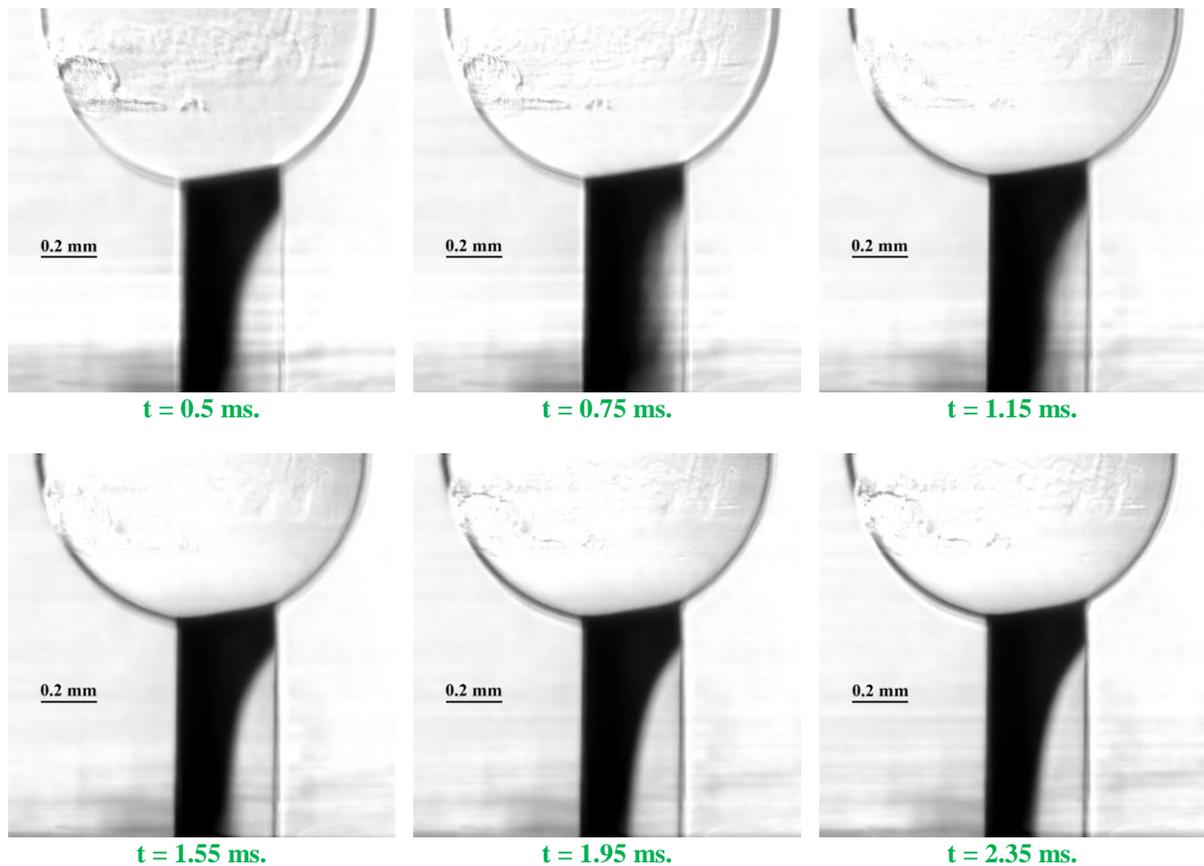

**Figure 5: Average cavitation images (sample size 25) for different injector needle lifts at 300 bar of injection pressure.**

Spray cone angles were estimated using the spray images recorded using the two orthogonal cameras (Cam 1 and Cam 2 in Figure 2) from the average spray images (sample size 200 images) by converting them to binary images using a threshold value. The binary image was then used to calculate the spray cone angle by fitting the two side edges of the spray with straight lines as shown on the left of Figure 7. The angles between these straight lines and the centerline (nozzle orifice axis) are plotted individually in Figure 7. It can be observed that the spray images recorded through Cam 1 are almost symmetric (left and right angles are almost the same), whereas a significant difference between the left and the right side angles is observed for the spray images recorded through Cam 2. It is a consequence of the asymmetric nozzle design. Cavitation is more developed on one side of the nozzle clearly visible via Cam 2. However, for the orthogonal visualization axis the entire nozzle hole seems to be covered by the cavitation bubbles as shown in the cavitation images along the cameras in Figure 2 and hence the spray images are symmetric for this viewing angle.

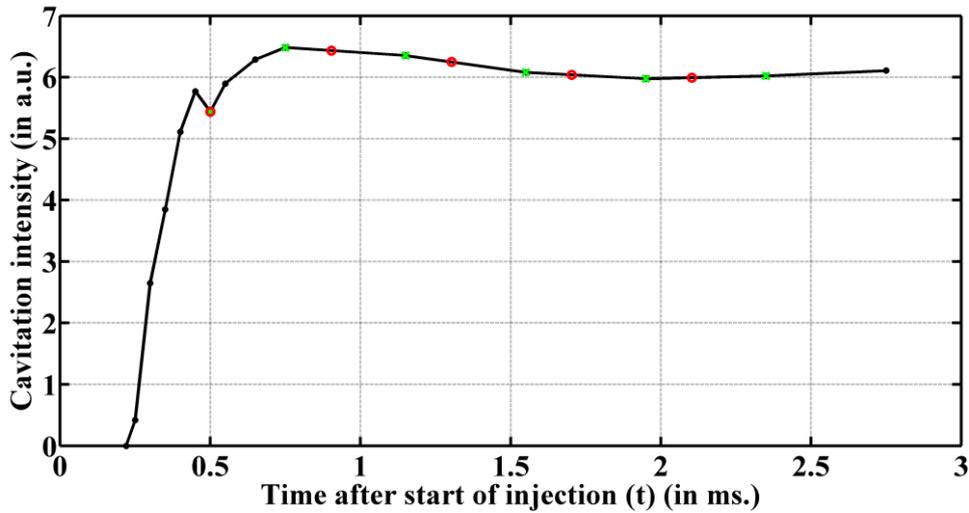

**Figure 6: Cavitation intensity in arbitrary units for different time after start of the fuel injection. The green and red marks indicate the time at which the cavitation and spray images were recoded respectively for comparison.**

From the cavitation images in Figure 5 and corresponding cavitation intensity plotted in Figure 6 it can be seen that the cavitation reaches a maximum at t = 0.75 ms after the start of the fuel injection (corresponding needle lift = 24.3%). Spray cone angle also seems to follow the same trend as the cavitation intensity, with first an increasing period lasting for about 1 ms. However, more such measurements for different needle lifts are required to ascertain this behavior.

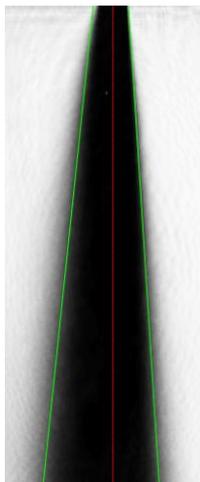 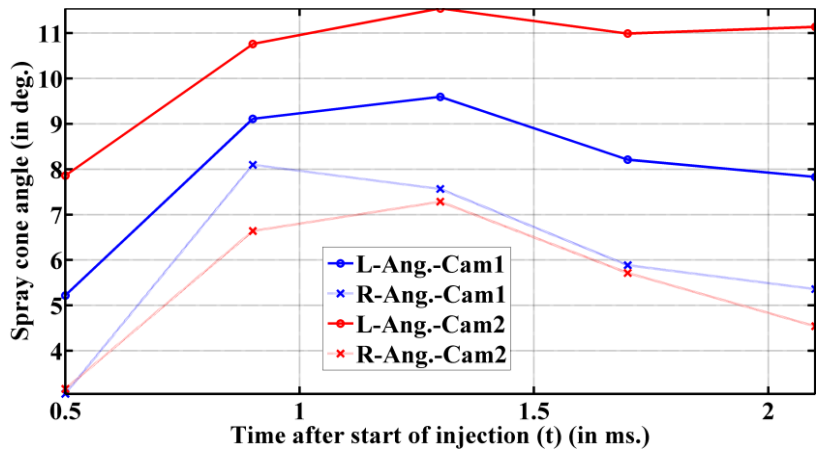

**Figure 7: On the left is an average of 200 spray images recorded with Cam 2 for t = 0.9 ms after start of injection and the fitted straight lines for spray cone angle estimation and the plot on the right shows its evolution with time after start of fuel injection for the two orthogonal views (L and R corresponds to the left and right sides of the spray respectively).**

In order to observe the overall behaviour of the spray velocity for different needle lifts and the role of cavitation on it in the near nozzle region, average velocity maps were computed from 200 image-pairs using the cross-correlation based approach. Figure 8 shows the velocity maps for the spray image-pairs recorded using Cam 1 for various time delays after start of the fuel injection mentioned in Table 2 (in red). These velocity maps are almost symmetric and follow a very similar trend as followed by the cavitation intensity for various time delays.

For the side view (images recorded with Cam 2), one would expect to see a clear asymmetrical behaviour in the velocity maps as well due to cavitation. The cavitation bubbles develop on the left side of the injector nozzle (see Figure 5) and the velocity of the spray is higher on the right than on the left side as can be seen for all the velocity maps in Figure 9. This difference in liquid velocities lead to different development of liquid atomization on each side of the jet. In case of atomisation controlled by a characteristic time scale, the liquid structures on the right side of the jet moving faster would undergo primary breakup at a larger distance from the nozzle outlet compared to those on the left side. As a consequence, atomization occurs at a shorter distance from the nozzle outlet in presence of cavitation. This is one of the benefit of cavitation in diesel combustion applications, particularly in the context of engine downsizing.

**Summary and Conclusions**

The aim of the transparent nozzle head design to have a stable cavitation on one side of the nozzle at high-pressures was successfully achieved using the PMMA asymmetric nozzle with orifice diameter of 0.35 mm. The transparent nozzle head was tested for an injection pressure of 300 bar. The average cavitation intensity was estimated for various needle lifts at this injection pressure from the average cavitation images. It was found that for our transparent nozzle the cavitation increases rapidly until 1/10th of the injector needle is lifted and then gets stabilized until it is completely lifted.

The spray characteristics were also estimated using the spray images obtained for two orthogonal views. A double pulsed laser system was used to obtain two spray images for a single injection within a short duration of time for velocity measurement. A large set of images (about 200 image-pairs) were obtained and used for average velocity map computation for five different needle lifts. It is observed that strong velocity variations are observed when cavitation evolves during the needle lift. These images were also used for spray cone angle computation and the obtained results are presented and compared with the cavitation intensity.

**Acknowledgements**

This work was supported by CANNEx program (ANR-13-TDMO-03), funded from French National Research Agency (ANR).

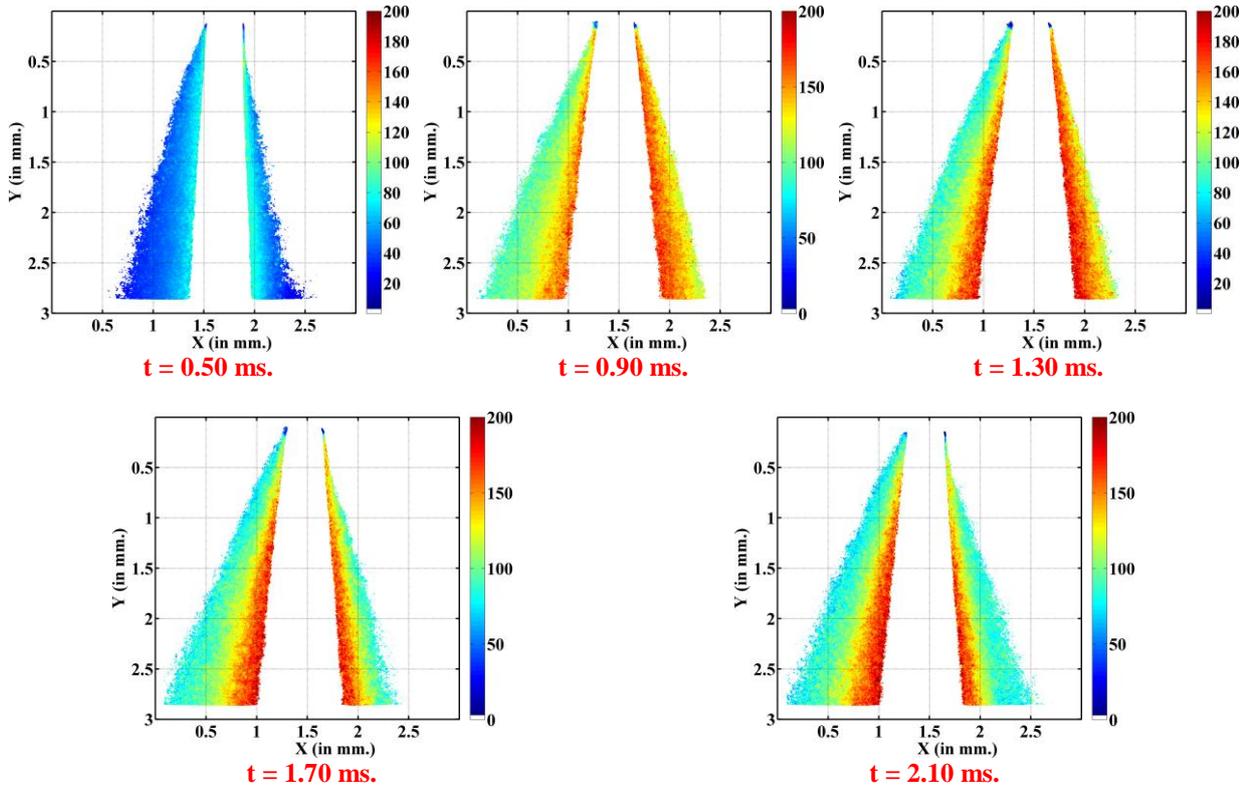

**Figure 8:** Spray velocity maps at different time after start of the fuel injection computed using the spray images recorded by Cam 1 of Figure 2.

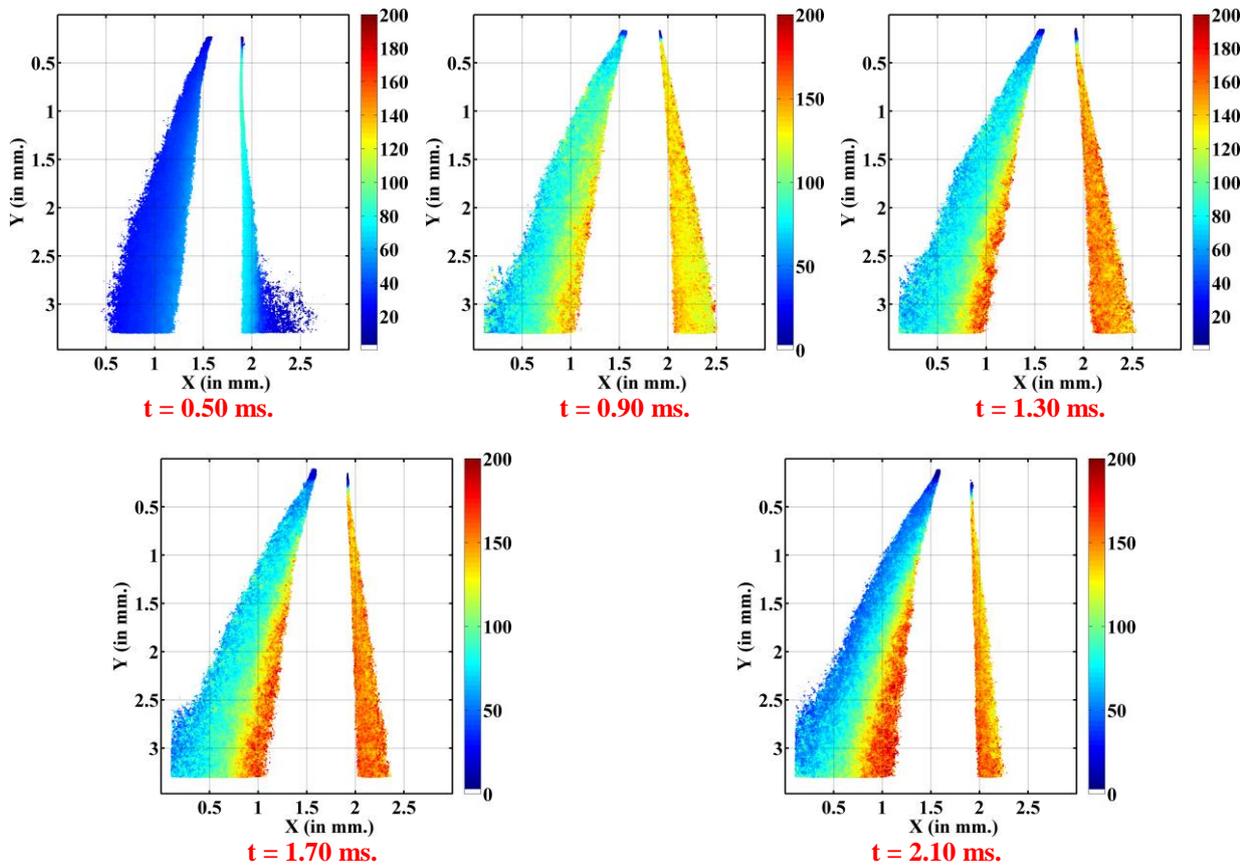

**Figure 9:** Spray velocity maps at different time after start of the fuel injection computed using the spray images recorded by Cam 2 of Figure 2.